\documentstyle[12pt]{article}
\newcommand {\beq}{\begin{equation}}
\newcommand {\eeq}{\end{equation}}
\newcommand {\beqa}{\begin{eqnarray}}
\newcommand {\eeqa}{\end{eqnarray}}
\newcommand {\beqan}{\begin{eqnarray*}}
\newcommand {\eeqan}{\end{eqnarray*}}
\newcommand {\n}{\nonumber \\}

\newcommand {\eq}[1]{eq.~(\ref{#1})}

\newcommand {\Romannumeral}[1]{\uppercase\expandafter{\romannumeral#1}}

\newcommand {\del}{\partial}

\begin{document}
\setlength{\oddsidemargin}{0cm}
\setlength{\baselineskip}{7mm}

\begin{titlepage}
 \renewcommand{\thefootnote}{\fnsymbol{footnote}}
    \begin{normalsize}
     \begin{flushright}
                     TIT-HEP-293\\
			HUTP-95/A018\\
                    May 1995~~
     \end{flushright}
    \end{normalsize}
    \begin{Large}
       \vspace{1cm}
       \begin{center}
         {\LARGE Quantum Gravity is Renormalizable\\
                 near Two Dimensions} \\
       \end{center}
    \end{Large}

  \vspace{10mm}

\begin{center}
           Yoshihisa K{\sc itazawa}\footnote
           {E-mail address : kitazawa@phys.titech.ac.jp
and kitazawa@huhepl.harvard.edu}\\
        {\it Department of Physics,
Tokyo Institute of Technology,} \\
            {\it Oh-okayama, Meguro-ku, Tokyo 152, Japan}
	\footnote{Permanent address and address after May 29}\\
	{\it and}\\
	{\it Lyman Laboratory of Physics,
Harvard University,}\\
	{\it Cambridge, MA 02138, USA}\\
\vspace{15mm}

\end{center}
\begin{abstract}
\noindent
We prove the renormalizability of quantum gravity
near two dimensions. The successful strategy is to keep
the volume preserving diffeomorphism as the manifest
symmetry of the theory.
The general covariance is recovered by
further imposing the
conformal invariance.
The proof utilizes BRS formalism
in parallel with Yang-Mills theory.
The crucial ingredient of the proof is the relation
between the conformal anomaly and the $\beta$ functions.
\end{abstract}
\end{titlepage}
\vfil\eject

\section{Introduction}

In two dimensions, Einstein-Hilbert action is
well known to be topological. Therefore the
Einstein tensor identically vanishes in two dimensions.
On the other hand, the $\beta$ function of the
gravitational coupling constant $G$ indicates that the
theory is asymptotically free\cite{2+epsilon}.
The $\beta$ function of the gravitational coupling
constant in $D=2+\epsilon$ dimensions at the one loop level
is\cite{KN,KKN1}
\beq
\mu{\del \over \del \mu} G
= {\epsilon G} - {25-c\over 24\pi}G^2,
\label{betag}
\eeq
where $\mu$ is a renormalization scale
and $c$ counts the matter contents.
It shows that the theory is well defined at short
distance as long as $c < 25$.
The short distance fixed point of the $\beta$
function is $G^*=24\pi\epsilon /(25-c)$.

We can take the two dimensional limit
of this fixed point if we let $c \rightarrow 25$
simultaneously
with arbitrary strength of $G^*$.
This is the critical string.
If $c<25$, the gravitational
coupling constant grows at long distance as
$8\pi/G \sim Q^2 log (\mu)$
in two dimensions
where $Q^2 = (25-c)/3$.
Therefore we need the cosmological constant
$\Lambda$
to define the theory.
The scaling dimension of the cosmological
constant is
$\Lambda \sim ({1/\mu})^{Q \alpha}$, where
$\alpha = {Q\over 2}(1-\sqrt{1-{8\over Q^2}})$
\cite{Polyakov,DDK,KKN1,KKN2}.
Hence we naturally explain the double scaling
relation of the matrix models\cite{Gross}:
\beq
exp(-8\pi/G) \sim \Lambda^{Q\over \alpha}.
\eeq
However the theory is not always controllable
in this way. For $1<c<25$,
$\alpha$ becomes complex and the infrared behavior
of the theory appears to be too wild.

Therefore somewhat mysterious
asymptotic freedom of the topological
gravitational coupling constant fits our knowledge of
string theory very nicely.
If we take the asymptotic freedom seriously, we may
contemplate the possibility that consistent
quantum gravity exists beyond two dimensions.
The $2+\epsilon$ dimensional expansion of quantum gravity
enables us to study such an idea systematically.
The pocket of a new phase with width $\epsilon$ opens up
for $0<G<G^*$ in quantum gravity
beyond two dimensions.
In this weak coupling phase physical theories
with $1<c<25$ are well defined since the infrared
behavior of the theory is trivial.
Furthermore this phase resembles our universe.

However the difficulty of this program
has been recognized in \cite{KKN1} which
may be rephrased in the following way.
Let us treat the conformal mode of the
metric as a matter field.
We also need to introduce a reference metric.
 From such a view point, the general covariant theory
always possesses the conformal invariance
with respect to the reference metric.
However it is also well known that it is not possible
to maintain the conformal invariance in quantum
field theory with nontrivial $\beta$ functions.
Therefore we are inescapable from the
presence of the anomaly
in the crucial symmetry of the theory.

It has been proposed to only keep the volume
preserving diffeomorphism as the manifest
symmetry. The general covariance can be
recovered by further imposing the conformal
invariance with respect to the reference metric
on the theory including the quantum
effect\cite{KKN2,AKKN}.
A possible proof of this proposal was suggested
in \cite{AKKN} and it was checked by an explicit calculation
at the two loop level\cite{AKNT}.
In this paper we prove the renormalizability
of quantum gravity near two dimensions through
this approach.
We thereby lay the foundation for the
$2+\epsilon$ dimensional expansion
of quantum gravity.
We hope that this approach also sheds light on the
lattice approach\cite{AM,AJ}.

In section two, we recall the formulation of the
quantum gravity in $2+\epsilon$ dimensions.
We set up the BRS formalism
and derive the Ward-Takahashi identities.
In section three, we solve the WT identity to determine
the bare action. We give the inductive proof of the
renormalizability in section four. We show that the
divergences of the theory can be canceled by the counter terms
which can be supplied by the bare action.
We conclude in section five with discussions.

\section{BRS Invariance and Ward-Takahashi Identity}

In this section, we recall the formulation of
quantum gravity in $2+\epsilon$ dimensions.
We then set up the BRS formalism and derive
Ward-Takahashi identities.
We decompose the metric into the conformal factor
and the rest as $g_{\mu\nu} = \hat{g}_{\mu\rho}
(e^h)^{\rho}_{~\nu}e^{-\phi} = \tilde{g}_{\mu\nu}e^{-\phi}$.
$\hat{g}_{\mu\nu}$ is a background metric.
The tensor indices of fields are raised and lowered by
the background metric. $h_{\mu\nu}$ is a traceless symmetric
tensor.
We have decomposed the metric into two different types
of variables since their renormalization properties
are very different\cite{KKN1,KKN2,AKKN}.

We consider the following generic action:
\beq
I = \int {\mu ^{\epsilon}\over G}
\{\tilde{R}\Phi(X) + {1\over 2} \tilde{g}^{\mu\nu}G_{ij}(X)
\del _{\mu} X^i \del _{\nu} X^j \} ,
\eeq
where $\int = \int d^Dx\sqrt{\hat{g}}$ denotes the
integration over the $D$ dimensional spacetime.
In this expression, $X^i$ are $N=c+1$ copies of real scalar
fields. The conformal mode of the metric is treated as one
of them. $\tilde{R}$ is the scalar curvature made out
of $\tilde{g}_{\mu\nu}$.
$G$ and the wave function
renormalization of $X^i$ fields are fixed by requiring
$\Phi(0)=1$ and $G_{ij}(0) = \eta_{ij}$
where $\eta_{ij}$ is the flat metric in
$N$ dimensions.
Generic theories with general covariance can be
described in this way by further imposing the
conformal invariance with respect to the background
metric.
Our approach is therefore
a natural generalization of the nonlinear sigma
model approach in string theory
\cite{CFMP,FT,AT} into higher
dimensions.

The crucial symmetry of the theory is the invariance
under the following gauge transformation:
\beqa
\delta \tilde{g}_{\mu\nu} & = &
\del _{\mu} \epsilon ^{\rho} \tilde{g} _{\rho\nu}
+\tilde{g}_{\mu\rho} \del _{\nu}\epsilon ^{\rho}
+ \epsilon ^{\rho} \del _{\rho} \tilde{g} _{\mu\nu}
-{2\over D} \del _{\rho} \epsilon ^{\rho}
\tilde{g} _{\mu\nu} ,\nonumber \\
\delta X^i & = & \epsilon ^{\rho} \del _{\rho}
X^i - (D-1) G^{ij} {\del \Phi \over \del X^j}
{2\over D} \del _{\rho} \epsilon ^{\rho} .
\label{gaugetr}
\eeqa
In order to prove the renormalizability of the theory,
we set up the BRS formalism.
The BRS transformation of these fields $\delta _B$
is defined by replacing the gauge parameter by the
ghost field
$\epsilon ^{\mu} \rightarrow C^{\mu}$.
The BRS transformation of $h_{\mu\nu}$ field
is defined through the relation $\tilde{g} =
\hat{g}e^h$.
The BRS transformation of ghost, antighost
and auxiliary field is
\beqa
\delta _B C^{\mu}& = & C^{\nu} \del _{\nu} C^{\mu},
\nonumber \\
\delta _B \bar{C}^{\mu} & = & \lambda ^{\mu},
\nonumber \\
\delta _B \lambda ^{\mu} & = & 0.
\label{brstr}
\eeqa
The BRS transformation can be shown to be
nilpotent $\delta _B^2 = 0$.

Our proof proceeds in parallel with Yang-Mills theory
case\cite{ZJ}. However the conformal anomaly
forces us further investigations.
We denote $A_i =
(h_{\mu\nu} , X_i)$. We also introduce a gauge fixing
function $F_{\alpha} (A)$.
It is an arbitrary function of $A$ with dimension one.
The gauge fixed action is
\beq
S = I +
\int
[- {G\over 2\mu^{\epsilon}} \lambda _{\alpha}\lambda ^{\alpha}
+\lambda ^{\alpha} F_{\alpha} - \bar{C}^{\alpha}
\delta _B F_{\alpha} - K^i \delta _B A_i
+L_{\alpha} \delta _B C^{\alpha} ].
\eeq
Here we have introduced sources $K$ and $L$ for the
composite operators.
The criterion for the action $I$ is
the invariance under the volume preserving
diffeomorphism.
Hence the BRS invariance is broken if there exits
conformal anomaly:
\beq
\delta_B I = \int  T^{\alpha}_{~\alpha}
{2\over D} \del_{\beta}C^{\beta},
\eeq
where $T^{\alpha}_{~\alpha} = - \hat{g}_{\mu\nu}
\delta I / \delta \hat{g}_{\mu\nu}$.

The partition function is
\beq
Z=e^W=\int [dAdCd\bar{C}d\lambda]
exp(-S+\int[J^iA_i+\bar{\eta}_{\alpha}C^{\alpha}
+\bar{C}^{\alpha}\eta_{\alpha} + l_{\alpha}\lambda^{\alpha}]).
\eeq
By the change of the variables with BRS transformation form,
we obtain the Ward-Takahashi identity for the generating
functional of the connected Green's functions:
\beq
\int (J^i{\delta \over \delta K^i} +
\bar{\eta}_{\alpha}{\delta \over \delta L_{\alpha}}
+\eta_{\alpha}{\delta \over \delta l_{\alpha}})W
= <\int T^{\alpha}_{~\alpha} {2\over D} \del_{\beta}
C^{\beta}>.
\label{wtw}
\eeq

The WT identity for the effective action is
obtained by the Legendre transformation:
\beq
\int[
{\delta \Gamma \over \delta A_i}
{\delta \Gamma \over \delta K^i} +
{\delta \Gamma \over \delta C^{\alpha}}
{\delta \Gamma \over \delta L_{\alpha}}
-\lambda ^{\alpha}
{\delta \Gamma \over \delta \bar{C}^{\alpha}}
]
=
-\int T^{\alpha}_{~\alpha}{2\over D}\del _{\beta}
C^{\beta} .
\label{wtpv}
\eeq
In order to make the above expression finite,
we need to add all necessary counter terms to
$S$.
The bare action $S^0$ obtained in this way
satisfies the same equation
\beq
\int [
{\delta S^0 \over \delta A_i}
{\delta S^0 \over \delta K^i} +
{\delta S^0 \over \delta C^{\alpha}}
{\delta S^0 \over \delta L_{\alpha}}
-\lambda ^{\alpha}
{\delta S^0 \over \delta \bar{C}^{\alpha}}]
=
-
\int T^{\alpha}_{~\alpha}{2\over D}\del _{\beta}
C^{\beta}.
\label{wts}
\eeq
On the other hand, \eq{wtw} follows from \eq{wts}
in dimensional regularization.
Therefore the following famous relation between the trace anomaly
and the bare action holds as an operator identity:
$
T^{\alpha}_{~\alpha} = - \hat{g}_{\mu\nu}
\delta I^0 /\delta \hat{g}_{\mu\nu}
$.
To simplify notations, we introduce an auxiliary field $M_{\alpha}$
and add to the action the combination $-\int M_{\alpha}\lambda^{\alpha}$
in such a way that $\lambda ^{\alpha}=-{\delta \Gamma\over \delta
M_{\alpha}} = -{\delta S\over \delta M_{\alpha}}$.
Then the left hand side of \eq{wtpv} and \eq{wts} become
homogeneous quadratic equations which we write symbolically
as $\Gamma *\Gamma$ and $S^0*S^0$ .

\section{Analysis of the Bare Action}

In this section, we solve \eq{wts} to determine $S^0$.
In the subsequent considerations, we deal with the bare fields
and the bare BRS transformation.
The bare BRS transformation is given in \eq{gaugetr} and \eq{brstr}
in terms of the bare fields.
The precise relation between the bare
and the renormalized quantities will be explained
in the next section.
$S^0$ can always be decomposed into the part $I^0$ which involves only
$A_i$ fields and the rest. The right hand side of \eq{wts}
is determined by $I^0$ only.

Let us examine the general structure of the bare action.
By power counting, it has to be a
local functional of fields and sources
with dimension $D$.
We also have the ghost number conservation rule
and its ghost number has to be zero.
By these dimension and ghost number considerations,
it is easy to see that
$K$ and $L$ appear only linearly in $S^0$:
\beq
S^0 = \int
[ - K^i(\delta_B' A_i) +L_{\alpha}(\delta_B' C^{\alpha})]
+\tilde{S},
\eeq
where $\delta_B'$ denotes most general BRS like
transformations
with correct dimension and ghost number.
It is also easy to see that there are no $\lambda$ and hence
no $\bar{C}$ dependence in $\delta_B'$.
%
%
Since $\lambda$ has dimension $1+\epsilon$, $\tilde{S}$ can be
at most quadratic in $\lambda$:
\beq
\tilde{S} = \int [
-{G^0\over 2} \tilde{E}_{\alpha\beta}\lambda^{\alpha}\lambda^{\beta}
+\lambda^{\alpha}\tilde{F}_{\alpha}
+\tilde{L}],
\eeq
where $\tilde{E}_{\alpha\beta}$ and $\tilde{F}_{\alpha}$ are
general functions of $A,C$ and $\bar{C}$
with dimension zero and one respectively.
$G^0$ is the bare gravitational coupling constant
and it is the only quantity with dimension $-\epsilon$.

In order to determine the structure of the bare action, we
decompose $S^0 = S + \delta S$ and $I^0 = I + \delta I$.
We may assume without loss of generality that
$\delta S$ and $\delta I$ are small.
General solutions may be obtained by integrating these
solutions.
Then we obtain the following equation
for $\delta S$:
\beq
\int \Delta \delta S = -\int T^{\alpha}_{~\alpha} {2\over D}
\del_{\beta}C^{\beta},
\label{delx}
\eeq
where the trace anomaly on the right hand side is that of
$\delta I$.
Here we have introduced a differential operator:
\beq
\Delta = {\del S\over \del A_i}{\del \over \del K^i}
+ {\del S\over \del K^i}{\del \over \del A_i}
+ {\del S\over \del C^{\alpha}}{\del \over \del L_{\alpha}}
+ {\del S\over \del L_{\alpha}}{\del \over \del C^{\alpha}}
+ {\del S\over \del M_{\alpha}}{\del \over \del \bar{C}^{\alpha}}.
\eeq
We denote below by $\theta^i$ the set of all anticommuting fields
$K^i,C^{\alpha},\bar{C}^{\alpha}$ and $x_i$ all
commuting fields $A_i,L_{\alpha},M_{\alpha}$.
Under the following infinitesimal change of the variables
\beqa
x_i'& = & x_i - {\del \psi\over \del \theta^i},\n
(\theta^i)' & = & \theta^i + {\del\psi\over \del x_i},
\eeqa
the action $S(\theta^i ,x_i)$ changes as
\beq
S(\theta',x') -S(\theta ,x) =  \Delta \psi.
\eeq
We also have the following relation:
\beq
\Delta ^2 = [-{\del\over \del\theta^j}
(S*S)]
{\del\over \del x_j} +
[{\del\over \del x_j}
(S*S)]{\del\over\del\theta^j}.
\eeq

The most general solution for $\delta S$ which satisfies
\eq{delx} is
\beq
\delta S=\int \delta_B (
\bar{C}^{\alpha}
(F'_{\alpha} + G^0\lambda^{\beta}E'_{\alpha\beta}))
+ \delta I(A),
\eeq
where $\delta I(A)$ is invariant under the
volume preserving diffeomorphism.
$E'$ and $F'$ are general functions of $A,C$ and $\bar{C}$
with dimension zero and one respectively.

The BRS exact part can be associated with a canonical
transformation on the fields.
It can be understood as a freedom in association with
the gauge fixing procedure.
Therefore we conclude that the bare action is
similar to the tree level action
in terms of the bare fields with
arbitrariness in association with the
gauge fixing procedure.
%

\section{Inductive Proof of the Renormalizability}

In this section, we give the inductive proof of the
renormalizability of quantum gravity by the $2+\epsilon$
dimensional expansion approach.
We will show that all necessary counter terms can be
supplied from the bare action
which is invariant under the volume preserving diffeomorphism.
We expand the trace anomaly
by the power series of $G$ as $T^{\alpha}_{~\alpha}
=\sum_{i=0}^{\infty} \beta _i G^i$. Our strategy is to tune it to be $O(G^l)$
at $l$ loop level.
Note that we have assumed it starts at $O(1)$.
We need to fine tune the tree action $I$ for this purpose
to start with since it is $O(1/G)$ in general.

Our analysis is based on a loopwise expansion of
the effective action:
\beq
\Gamma = \sum _{l=0}^{\infty} \Gamma _{l},
\eeq
in which $\Gamma _0$ is the tree level action S.
We assume as an induction hypothesis that we have been able to
construct an action $S^0_{l-1}$ with counter terms
which satisfies \eq{wts}
and renders $\Gamma$ finite at $l-1$ loop order.
Then the right hand side of \eq{wts} is
proportional to the trace anomaly.

Let the couplings and the operators dual to them as
$\{G,\lambda _k\}$ and $\{\tilde{R},\Lambda ^k\}$.
In our action, the couplings are $\{G,\Phi -1,G_{ij}-\eta_{ij}\}$.
$\lambda _k$ represents two arbitrary functions of
$X^i$ fields and hence is equivalent to the infinite numbers of
coupling constants.
The bare couplings are
\beqa
1\over G^0 & = &
{\mu^{\epsilon}\over G} Z_G
=\mu ^{\epsilon} ({1\over G} - \sum _{\nu =1}
^{l-1} {a^{\nu}_0\over \epsilon ^\nu}),\n
\lambda^{0}_{k} & = &
\lambda _k + \sum _{\nu =1}^{l-1}
{a^{\nu}_k\over \epsilon ^{\nu}}.
\eeqa
Here we have $1/\epsilon ^{l-1}$ poles at most
at $l-1$ loop order.
The $\beta$ functions follow as
\beqa
\mu{\del\over \del \mu}{G} & = & \epsilon G +\beta _G\n
& = &
\epsilon G - a^1_0 G^2 - G^3{\del\over \del G}
a^1_0 ,\n
\mu{\del \over \del \mu} \lambda _k & = & \beta _{\lambda_k}
 =  -G{\del\over \del G} a^1_k .
\label{betaf}
\eeqa
As it is well known, only the residues of the simple
pole in $\epsilon$ contribute to the $\beta$ functions. The coefficients
of the higher poles in the bare couplings
are determined by the finiteness of the
$\beta$ functions (pole identity).

The bare action is
\beq
I^0 = \int [{1\over G^0}\tilde{R}^0
+{1\over 2G^0}\eta_{ij}\del_{\mu}X_0^i\del_{\nu}X_0^j
\tilde{g}^{\mu\nu}
+{\lambda_k^0\over G^0}\Lambda ^k_0].
\eeq
The trace anomaly of the bare action is
\beqa
&&{1\over G^0}(\epsilon \Phi ^0 + 2(D-1)\del^i\Phi ^0\del _i
\Phi ^0) \tilde{R}\n
&+&{1\over 2G^0}(\epsilon G^0_{ij} + 4(D-1) \nabla _i\del _j\Phi ^0)
\del _{\mu}X^i_0\del _{\nu}X^j_0\tilde{g}^{\mu\nu}.
\label{bta}
\eeqa
We need to rewrite this expression in terms of the
renormalized operators\cite{AT}.
The wave function renormalization can be ignored
since we are investigating the renormalization
of the operators after the wave function
renormalization.
It can be justified by using the equations of motion
in dimensional regularization.
By the same reason we can do away with the total
derivatives in the trace anomaly.

We have shown that the bare action consists of
the BRS exact part and the part which is invariant
under the volume preserving diffeomorphism.
We assume as an inductive hypothesis
that we have been able to renormalize
the theory up to $l-1$ loop level with counter terms
which can be supplied by the bare action.
Namely the counter terms themselves consist of
the BRS exact part and the part invariant under the
volume preserving diffeomorphism.
Therefore we assume that the operator mixing has
occurred only within each sector,
namely the BRS exact operators and the
operators which are invariant under the
volume preserving diffeomorphism.
For this reason we can ignore the BRS exact
operators in the following considerations.

We introduce the renormalized operators which are
defined as
\beqa
\tilde{R} & = & (-G^2 {\del\over \del G}
-G\lambda_k{\del\over \del\lambda_k})I^0,\n
\Lambda_k & = & G{\del\over \del\lambda_k}I^0.
\label{fno}
\eeqa
This operator is finite up to $l-1$
loop order by our inductive assumption.
We fix the wave function renormalization
of $X^i$ fields
as ${1\over G^0}\eta_{ij}\del_{\mu}X^i_0
\del_{\nu}X^j_0\tilde{g}^{\mu\nu} = {\mu^{\epsilon}\over G}
\eta_{ij}\del_{\mu}Y^i\del_{\nu}Y^j\tilde{g}^{\mu\nu}$.
For this purpose
we rewrite $\sqrt{Z_G}X^i_0
=Y^i$ and do not differentiate $Y^i$
in \eq{fno}.
We also subtract the already finite kinetic term
for $X^i$ fields from $\tilde{R}$.

We can expand the trace anomaly \eq{bta} in terms
of the renormalized couplings and operators.
The trace anomaly
is
\beqa
&&{1\over G}\{\epsilon \Phi +
{\beta _G\over G} -\beta _{\Phi} +
{\beta_G\over G}(1-{1\over 2} X^i{\del\over \del X^i})
(\Phi -1)
+2(D-1)\del^i\Phi\del_i\Phi\}
\tilde{R}
\n
&+&{1\over 2G}\{\epsilon G_{ij}
-\beta _{G_{ij}}
-{\beta _G\over G} {1\over 2}{X^k{\del\over \del X^k}}
G_{ij}
+4(D-1)\nabla _i\del_j\Phi\}
\del_{\mu}X^i\del_{\nu}X^j\tilde{g}^{\mu\nu}.
\label{trace}
\eeqa

In this way, the trace anomaly
can be expressed by the $\beta$ functions and
the renormalized operators.
The singularities in $\epsilon$ cancel out
up to $O(G^{l-2})$ by the assumption.
We can tune the couplings
in the theory such that the conformal anomaly vanishes
up to $O(G^{l-2})$.
Now we have tuned the coupling constants of the theory
such that the
the right hand side of
\eq{wts} is $O(G^{l-1})$.
At this order, the trace anomaly has divergences
in general.
However these divergences are
the inevitable consequences of the
lower loop divergences.
Let us add the counter terms at $l$ loop order
which are required by the pole identity to make
$\beta$ functions finite. These counter terms
with higher poles in $\epsilon$ are
invariant under the volume preserving diffeomorphism.
In this way we can make the right hand side of
\eq{wts} finite at $l$ loop order.

Now we can solve \eq{wtpv} at $O(G^{l-1})$
to obtain:
\beq
S*\Gamma _l + \Gamma _l *S = \Delta \Gamma_l
= -\sum _{m=1}^{l-1}
\Gamma _m * \Gamma _{l-m} - T^{\alpha}_{~\alpha}{2\over D}
\del _{\beta}C^{\beta}.
\label{s*g}
\eeq
This equation determines the divergent part of
$\Gamma _l^{div}$.
Since we have subtracted
all subdivergences, the divergences are local.
The solution of this equation can be decomposed
into the BRS exact part and the rest.
The nontrivial part of
$\Gamma _l^{div}$
has to be invariant under the volume preserving
diffeomorphism.
Furthermore new divergences
which contribute to the $\beta$ functions at $l$ loop
order have to be conformally invariant
in two dimensional limit.
It is because the right hand side of \eq{s*g}
has been made finite.

$\Gamma _l^{div}$
may also have the BRS exact part of $\Delta\psi$ form
since $\Delta^2 = 0$ to this order.
The general form of $\psi$ is:
\beq
\psi =
K^i\Psi '_i +L_{\alpha}\Theta '^{\alpha}_{~\beta}C^{\beta} +
\bar{C}^{\alpha}
(F'_{\alpha} + G\mu^{-\epsilon}\lambda^{\beta}E'_{\alpha\beta}),
\eeq
where $\Psi '$ and $\Theta '$ are general functions of
$A,C,\bar{C}$ with dimension zero and vanishing ghost number.
As we have explained, the BRS exact part can be associated with
a canonical transformation on the fields.
Here we consider the physical implications of these
canonical transformations.
Under this transformation,
the part of $S^0$ linear in $K$ and $L$ changes as:
\beqa
K^i\delta_BA_i &\rightarrow& K^i\delta_BA_i
+K^i\delta_B(\Psi '_i)
-K^i{\del \delta_BA_i\over \del A_j}
\Psi '_j
+K^i{\del \delta_BA_i\over \del C^{\alpha}}
\Theta'^{\alpha}_{~\beta}C^{\beta},\n
L_{\alpha}\delta_BC^{\alpha} &\rightarrow&
L_{\alpha}\delta_BC^{\alpha} -
L_{\alpha}\delta_B(\Theta '^{\alpha}_{~\beta}C^{\beta})
+L_{\alpha}{\del \delta_BC^{\alpha}\over
\del C^{\beta}}\Theta '^{\beta}_{~\gamma}C^{\gamma}
-L_{\alpha}{\del \delta_BC^{\alpha}\over \del A_i}
\Psi '_i.
\eeqa

These infinitesimal deformations can be
interpreted as the change of the functional form of
the BRS transformation in association with
the wave function renormalization of the fields.
Note that
the functional form
of the BRS transformation has to change
in terms of the renormalized variables,
although
the functional form of the BRS transformation
remains the same in terms of the bare fields.
The renormalized BRS transformation continues to be nilpotent.
The rest of the BRS exact part causes the renormalization
of the gauge fixing part.

By defining the bare action at $l$ loop level
\beq
S^0_l=S^0_{l-1} - \Gamma _l^{div} +
higher~orders ,
\eeq
it is possible to render $\Gamma$ $l$ loop finite.
Here $\Gamma _l^{div}$ includes divergences with
higher poles as well as simple poles in $\epsilon$.
The counter terms can be interpreted as the coupling
constant and wave function renormalization
of a bare action as we have explained.
By doing so,
we are able to construct the bare action
$S_l^0$ which satisfies \eq{wts}.
Now the circle is complete and we have proven
the renormalizability
of quantum gravity near two dimensions.

Let us consider the following model
(conformal Einstein gravity)
as a concrete example\cite{AKKN,AKNT}:
\beqa
\Phi &=& 1+\sqrt{\epsilon}a\psi +\epsilon b(\psi^2
-\varphi^2_i),\n
G_{ij}& =& \eta_{ij},
\label{egpr}
\eeqa
where $\eta_{ij}$ is the flat Minkowski metric
in $N$ dimensions.
In this model, the operators which are
invariant under the volume preserving diffeomorphism and
conformally
invariant in two dimensions are
$\int \tilde{R},\int [\sqrt{\epsilon}a\psi\tilde{R}
-{1\over 2}\del_{\mu}
\psi\del_{\nu}\psi\tilde{g}^{\mu\nu}]$
and $\int {1\over 2}\del_{\mu}
\varphi_i\del_{\nu}\varphi_i\tilde{g}^{\mu\nu}$.
These operators are invariant under the
transformation \eq{gaugetr} up to $O(\epsilon )$.

Therefore at $l$ loop level, new divergences
of the following form may arise:
\beq
{\mu^{\epsilon}\over G}\int[
{a^1 G\over \epsilon}\tilde{R}
-{z^1\over \epsilon }
(\sqrt{\epsilon}a\psi\tilde{R}
-{1\over 2}\del_{\mu}
\psi\del_{\nu}\psi\tilde{g}^{\mu\nu})
+{z'\over \epsilon }
({1\over 2}\del_{\mu}
\varphi_i\del_{\nu}\varphi_i\tilde{g}^{\mu\nu})].
\eeq
The last divergence in the above expression can be
taken care of by the wave function renormalization of
$\varphi_i$.
The second part of these divergences can be dealt with
by the following wave function renormalization:
$\psi \rightarrow (1+{z^1\over 2\epsilon})\psi +{a
\over 4\sqrt{\epsilon}b}{z^1\over \epsilon}$.
By these wave function renormalizations,
we can supply the required counter terms
from the original action.
Since $b$ is associated with an explicit factor of
$\epsilon$, this procedure introduces a finite counter term
of $\mu^{\epsilon}\int (\psi^2-\varphi_i^2)\tilde{R}$ type.
The $\beta$ function of $b$
receives no contribution.
In order to cancel the remaining divergence,
we need the counter term
$-\mu^{\epsilon}\int (a_{0}^1/\epsilon )\tilde{R}$ where
$a_{0}^1 = a^1+z^1a^2/4bG$.

Therefore this model is renormalizable
to all orders
with the following bare action:
\beq
{\mu^{\epsilon}\over G}\int
[Z_G \tilde{R} +
\sqrt{\epsilon}a\psi\tilde{R}
+\epsilon b (\psi^2-\varphi_i^2)\tilde{R}
-{1\over 2}\del_{\mu}\psi\del_{\nu}\psi\tilde{g}^{\mu\nu}
+{1\over 2}\del_{\mu}\varphi_i\del_{\nu}\varphi_i\tilde{g}^{\mu\nu}].
\label{bareeg}
\eeq
The $\beta$ functions of $b$ vanishes
while $\mu {\del\over \del \mu}G=\epsilon GZ_G
/(1-G{\del\over \del G})Z_G$ agrees with \eq{betaf}.
The trace anomaly \eq{trace} becomes
\beqa
&&{1\over G}\{\epsilon Z_G
/(1-G{\del\over \del G})Z_G
+\epsilon(\Phi -1) +2(D-1)\del^i\Phi\del_i\Phi\}
\tilde{R}\n
&+& {1\over 2G}\{\epsilon \eta_{ij} + 4(D-1)\del_i\del_j
\Phi\}\del_{\mu}X^i\del_{\nu}X^j\tilde{g}^{\mu\nu}
\eeqa
It is finite as long as the $\beta$ function of $G$ is finite.
The trace anomaly vanishes to all orders if
\beqa
b&=&{1\over 8(D-1)},\n
\epsilon G +\beta _G & =& 2(D-1)a^2G\epsilon .
\label{tracecn}
\eeqa

The short distance fixed point of the renormalization group
where $a=\epsilon G +\beta_G=0$ is
certainly consistent with \eq{tracecn}.
Since the Einstein
action is recovered for large $\psi$, it may
be sufficient to construct the theory at the fixed point.
Here we draw the analogy with the spontaneous
symmetry breaking in field theory.
The theory with nonvanishing $a$ can be obtained
from the fixed point theory by
giving the vacuum expectation values to $\psi
\rightarrow \psi + a/(2\sqrt{\epsilon}b)$.
$a$ can be determined by the
second equation of \eq{tracecn}.
The renormalization group evolution
may be viewed as such a symmetry breaking
process and the renormalizability of the
theory on the whole renormalization group
trajectory naturally follows in such an interpretation.

\section{Conclusions and Discussions}

In this paper we have constructed a proof of the
renormalizability of quantum gravity near two
dimensions.
We thereby lay the foundation for the $2+\epsilon$
dimensional expansion of quantum gravity.
We have proven that all necessary counter terms can be
supplied by the bare action which is invariant
under the volume preserving diffeomorphism.
We can systematically cancel the trace anomaly
to all orders by tuning the coupling constants
of the theory.
In particular the Einstein gravity with
conformally coupled scalar fields is shown
to be renormalizable by tuning
the gravitational coupling constant to all
orders.

In this proof we have assumed that
the dimensional regularization preserves all
important symmetries of the theory. The Jacobian
in association with the change of the variables
which involves no derivatives is
always trivial in dimensional regularization.
We also need to assume some infrared
regularization.
A gauge invariant regularization is to consider
a closed finite universe.
However the simplest possibility is
to add a mass term to $h_{\mu\nu}$ field\cite{AKNT}.
Although it breaks the BRS invariance,
we can show that such a
soft breaking of the symmetry will not spoil
the renormalizability of the theory.

We need to modify the action in the following way:
\beq
S \rightarrow S + \int [{\mu^{\epsilon}\over 4G}m^2h_{\mu\nu}h^{\mu\nu}
+m^2 \bar{C}_{\alpha}C^{\alpha}
 + \tilde{M}\delta_B({\mu^{\epsilon}\over 4G}h_{\mu\nu}h^{\mu\nu}
+\bar{C}_{\alpha}C^{\alpha})],
\eeq
where $\tilde{M}$ is an external source.
The additional BRS exact part can be absorbed by the original action
by the canonical transformation of the external sources.
The WT identity \eq{wtpv}
for the effective action is modified as follows:
\beq
\Gamma *\Gamma = -T^{\alpha}_{~\alpha} {2\over D} \del_{\beta}C^{\beta}
-m^2{\del \Gamma \over \del \tilde{M}}.
\eeq
The \eq{s*g} which determines the divergence of $\Gamma _l$
is also modified as
\beq
S*\Gamma _l + \Gamma _l *S = \Delta \Gamma_l
= -\sum _{m=1}^{l-1}
\Gamma _m * \Gamma _{l-m} - T^{\alpha}_{~\alpha}{2\over D}
\del _{\beta}C^{\beta}
-m^2{\del \Gamma_l \over \del \tilde{M}}.
\eeq
The general divergence structure which is allowed by this
equation is
\beq
{\mu^{\epsilon}\over G}\int [m^2f(A,C,\bar{C})+\tilde{M}\delta_B f]
+\tilde{\Gamma}_l^{div},
\eeq
where $\tilde{\Gamma}_l^{div}$ is
obtained from $\Gamma_l^{div}$ in section four
by the canonical transformation of the
sources.
$f$ is a general function of $A,C$ and $\bar{C}$
with dimension zero.
Through this analysis
we have shown that the renormalization
property of the dimension two operators remains intact
while the soft symmetry breaking term may be renormalized
into a general dimension zero function.

Finally we make a comment on the Unitarity of the theory.
Let us assume that we are in the weak coupling phase.
The only poles in the Green's functions arise at
$p^2 = 0$ in the gravity sector.
However the theory flows to classical Einstein gravity
at long distance and
hence there should be no problem with
Unitarity in the theory.

\vspace{10mm}

I am grateful to E. D'Hoker, D. Gross and C. Vafa
for their hospitality at UCLA, Princeton and Harvard
University respectively
where part of this work has been carried out.
I also appreciate discussions on this subject
with H. Kawai, M. Ninomiya, T. Aida, J. Nishimura,
A. Tsuchiya and A. Migdal.
This work is supported in part by the Grant-in-Aid for Scientific Research
from the Ministry of Education, Science and Culture.

\newpage
\setlength{\baselineskip}{7mm}

\end{document}